\documentclass[aps,prd,12pt,preprint,tightenlines,superscriptaddress,
amsfonts,amssymb,amsmath,byrevtex,showpacs,nofootinbib]{revtex4}
\usepackage{graphics}
\usepackage{epsfig,subfigure}
\newcounter{mnotecount}[section]
\renewcommand{\themnotecount}{\thesection.\arabic{mnotecount}}
\newcommand{\mnote}[1]
{\protect{\stepcounter{mnotecount}}$^{\mbox{\footnotesize  $
      \bullet$\themnotecount}}$ \marginpar{\raggedright\tiny
    $\!\!\!\!\!\!\,\bullet$\themnotecount: #1} }

\begin{document}
\newcommand{\dR}{\mathbb R}
\newcommand{\dC}{\mathbb C}
\newcommand{\dZ}{\mathbb Z}
\newcommand{\id}{\mathbb I}

\title{The minimum length problem of loop quantum cosmology}

\author{Piotr Dzier\.{z}ak$^\dag$, Jacek Jezierski$^\ddag$, Przemys{\l}aw
Ma{\l}kiewicz$^\dag$, and
 W{\l}odzimierz Piechocki$^\dag$
\\ $^\dag$ Theoretical Physics Department, Institute for Nuclear Studies,
Ho\.{z}a 69, 00-681 Warsaw, Poland \\ $^\ddag$Department of
Mathematical Methods in Physics, University of Warsaw, Ho\.{z}a
69, 00-681 Warsaw, Poland}


\date{\today}

\begin{abstract}
The appearance of the big bounce (BB) in the evolution of the
universe is analyzed in the setting of loop quantum  cosmology
(LQC). Making use of an idea of a minimum length turns classical
Big Bang into BB. We argue why the spectrum of the kinematical
area operator of loop quantum  gravity  cannot be used for the
determination of this length. We find that the fundamental length,
at the present stage of development of LQC, is a free parameter of
this model.
\end{abstract}
 \maketitle

\section{Introduction}
Observational cosmology strongly suggests that our universe
emerged from a state with extremely high energy densities of
physical fields, called the  initial big-bang {\it singularity}.
Most of all models of the universe  obtained within the general
relativity (GR) also predict  the initial singularity
\cite{MTW,SWH,JPAK,JMMS}. It is commonly believed that the
singularity may be understood in a theory which unifies gravity
and quantum physics. Recent analysis done within the loop quantum
cosmology (LQC) concerning homogeneous isotropic universes of the
Friedmann-Robertson-Walker (FRW) type, strongly suggest that the
evolution of these universes does not suffer from the classical
singularity: the {\it big-bang} is replaced by  {\it big-bounce}
(with finite energy density of matter) owing to strong quantum
effects at the Planck scale
\cite{Ashtekar:2007em,Ashtekar:2006wn,Ashtekar:2006uz,Ashtekar:2003hd,
Bojowald:2006da}.

The goal of this paper is the {\it revision} of the foundation of
LQC concerning   the {\it minimum length}, $\mu_o$, which is
responsible for the resolution of the cosmological singularity. We
would like to attract an attention of the LQC community to the
problem of the {\it determination} of $\mu_o$. It has basic
meaning since its numerical value specifies the energy scale of
the Big Bounce transition. At the present stage of development of
LQC  the minimum length is a {\it free parameter}.

For simplicity of exposition we restrict ourselves to  the
quantization problem of the flat FRW model with massless scalar
field. This model of the universe unavoidably  includes the
initial cosmological singularity and has been  intensively studied
recently within LQC.

\section{Hamiltonian}
The gravitational part of the classical Hamiltonian, $H_g$, of GR
is a linear combination of the first-class constraints, and reads
\cite{TT,CR,Ashtekar:2004eh}
\begin{equation}\label{ham1}
    H_g:= \int_\Sigma d^3 x (N^i C_i + N^a C_a + N C),
\end{equation}
where $\Sigma$ is the space-like part of spacetime $\dR \times
\Sigma$, $~(N^i, N^a, N)$ denote Lagrange multipliers, $(C_i, C_a,
C)$ are the Gauss, diffeomorphism and scalar constraint functions.
In our notation  $(a,b = 1,2,3)$ are spatial and $(i,j,k = 1,2,3)$
internal $SU(2)$ indices. The constraint functions must satisfy a
specific algebra. It is known that this algebra (for constraints
smeared with test functions) is not a Lie algebra, but a Poisson
algebra because it includes structure functions instead of
structure constants (see, e.g. \cite{TT}).

In the case of flat FRW type universe with massless scalar field,
and with fixed local gauge and diffeomorphism freedom, the
classical Hamiltonian reduces to the scalar constraint and can be
shown (see, e.g. \cite{Ashtekar:2006uz}) to be

\begin{equation}\label{hamG}
H_g = - \gamma^{-2} \int_{\mathcal V} d^3 x
~~e^{-1}\varepsilon_{ijk}
 E^{aj}E^{bk} F^i_{ab},
\end{equation}
where  $\gamma$ is the Barbero-Immirzi parameter, $\mathcal
V\subset \Sigma$  is an elementary cell\footnote{In the case
$\Sigma$ is a non-compact manifold one introduces compact
submanifold $\mathcal V$ to give precise mathematical meaning of
the integrals.}, $~e:= \sqrt{|det E|}$, $~\varepsilon_{ijk}$ is
the alternating tensor, $~E^a_i $ is a densitised  vector field,
and where $F^i_{ab}$ is the curvature of an $SU(2)$ connection
$A^i_a$.

The resolution of the singularity, obtained within LQC, is based
on rewriting  the curvature $F^k_{ab}$ in terms of holonomies
around loops.   The curvature $F^k_{ab}$ can be determined
\cite{Ashtekar:2006uz} by making use of the formula
\begin{equation}\label{cur}
F^k_{ab}= -2~\lim_{Ar\,\Box_{ij}\,\rightarrow \,0}
Tr\;\Big(\frac{h^{(\lambda)}_{\Box_{ij}}-1}{\lambda^2
V_o^{2/3}}\Big)\;{\tau^k}\; ^o\omega^i_a  \; ^o\omega^j_a ,
\end{equation}
where
\begin{equation}\label{box}
h^{(\lambda)}_{\Box_{ij}} = h^{(\lambda)}_i h^{(\lambda)}_j
(h^{(\lambda)}_i)^{-1} (h^{(\lambda)}_j)^{-1}
\end{equation}
is the holonomy of the gravitational connection around the square
loop $\Box_{ij}$ which edges are parallel to the $i$- and
$j$-directions and of coordinate  length  $\lambda V_o^{1/3}$ with
respect to the flat fiducial metric $^o q_{ab}:= \delta_{ij}\, ^o
\omega^i_a\, ^o \omega^j_a $; fiducial triad $^o e^a_k$ and
co-triad $^o \omega^k_a$ satisfy $^o \omega^i_a\,^o e^a_j =
\delta^i_j$; spatial part of FRW metric is $q_{ab}=a^2(t)\,^o
q_{ab}$; $~Ar\,\Box_{ij}$ denotes the area of the square; $V_o =
\int_{\mathcal V} \sqrt{^o q} d^3 x$ is the fiducial volume of
$\mathcal V$; in what follows we set $V_o =1$ as its value is not
essential for our analysis.

The holonomy along straight edge of length $\lambda$ in the
$k$-direction (in the $j=1/2$ representation of $SU(2)$) may be
found \cite{Ashtekar:2006uz} to be
\begin{equation}\label{hol}
h^{(\lambda)}_k (c) = \cos (\lambda c/2)\;\id + 2\,\sin (\lambda
c/2)\;\tau_k,
\end{equation}
where $\tau_k = -i \sigma_k/2\;$ ($\sigma_k$ are the Pauli spin
matrices). It is clear that matrix elements of (\ref{hol}) can be
rewritten in terms of  $\exp (i \lambda c/2)$ which we denote by
$N_\lambda (c)$.

In what follows we  apply the `old' quantization scheme
\cite{Ashtekar:2006uz}, despite the fact that the `improved'
scheme \cite{Ashtekar:2006wn} is commonly used by  LQC community.
The reason is that mathematics underlying the old scheme has been
presented {\it clearly} in a comprehensive paper
\cite{Ashtekar:2003hd}. However, our results concern  both
methods.

One can show \cite{Ashtekar:2006uz} that $H_g$ may be rewritten as
\begin{equation}\label{hamR}
    H_g = \lim_{\lambda\rightarrow \,0}\; H^{(\lambda)}_g ,
\end{equation}
where
\begin{equation}\label{hamL}
H^{(\lambda)}_g = - \frac{sgn(p)}{2\pi G \gamma^3 \lambda^3}
\sum_{ijk} \varepsilon^{ijk}\, Tr \Big(h^{(\lambda)}_i
h^{(\lambda)}_j (h^{(\lambda)}_i)^{-1} (h^{(\lambda)}_j)^{-1}
h_k^{(\lambda)}\{(h_k^{(\lambda)})^{-1},V\}\Big),
\end{equation}
and where $V= |p|^{\frac{3}{2}}$ is the volume of the elementary
cell $\mathcal{V}$. The conjugate variables $c$ and $p$ satisfy
$\{c,p\} = 8\pi G \gamma /3$.  They determine {\it connections}
$A^k_a$ and density weighted {\it triads} $E^a_k$ due to the
relations $A^k_a = \,^o\omega^k_a\,c $ and $E^a_k =
\,^oe^a_k\,\sqrt{q_o}\,p $. However, $c$ and $p$ are not
elementary variables in (\ref{hamL}). The elementary functions
(variables) are chosen to be {\it holonomies} (described in terms
of $N_\mu$) and {\it fluxes} (proportional to $p$).

The classical total Hamiltonian for FRW universe with a massless
scalar field, $\phi$, reads
\begin{equation}\label{ham}
   H = H_g + H_\phi = 0,
\end{equation}
where $H_g$ is defined by (\ref{hamR}). The Hamiltonian of the
scalar field  is known to be: $H_\phi =  p^2_\phi
|p|^{-\frac{3}{2}}$, where $\phi$ and $p_\phi$ are the elementary
variables satisfying $\{\phi,p_\phi\} = 1$. The relation $H = 0$
defines the {\it physical} phase space of considered gravitational
system with constraints.

\section{Quantization}

In the Dirac quantization \cite{PAM,HT}  we find  a kernel of the
quantum operator $\hat{H}$ corresponding to $H$, i.e.
\begin{equation}\label{ker}
    \hat{H}\Psi = 0 ,
\end{equation}
(since the classical Hamiltonian is a constraint of the system),
and try to define a scalar product on the space of solutions to
(\ref{ker}). This gives a  starting point for the determination of
the physical Hilbert space $\mathcal{H}_{phys}$.

\subsection{Kinematics}

 The classical elementary functions satisfy the relation
\begin{equation}\label{relA}
\{p, N_\lambda \} = -i \frac{ 4\pi G  \gamma}{3} \lambda
N_\lambda,
\end{equation}
where $G$ is the Newton constant. Quantization of the algebra
(\ref{relA}) is done by making use of the prescription
\begin{equation}\label{pres}
    \{\cdot,\cdot\} \longrightarrow \frac{1}{i
    \hbar}\,[\cdot,\cdot].
\end{equation}
The basis of the representation space is chosen to be the set of
eigenvectors of the momentum operator \cite{Ashtekar:2003hd} and
is defined by
\begin{equation}\label{relB}
\hat{p}\,|\mu\rangle = \frac{4\pi\gamma l_p^2}{3}\, \mu
\,|\mu\rangle,~~~~\mu \in \dR ,
\end{equation}
where $l_p^2 = G \hbar$. The operator corresponding to $N_\lambda$
acts as follows
\begin{equation}\label{relC}
\hat{N}_\lambda  \,|\mu\rangle = |\mu + \lambda \rangle.
\end{equation}
The quantum algebra corresponding to (\ref{relA}) reads
\begin{equation}\label{relD}
    \frac{1}{i \hbar}[\hat{p},\hat{N}_\lambda ]\,|\mu\rangle = -i \frac{4\pi G
    \gamma}{3}\,\lambda\,\hat{N}_\lambda\, |\mu\rangle.
\end{equation}
The carrier space, $\mathcal{F}_g$, of the representation
(\ref{relD}) is the space spanned by $\{|\mu\rangle,\,\mu\in
\dR\}$ with the scalar product defined as
\begin{equation}\label{scal}
\langle \mu |\mu ^\prime \rangle:=  \delta_{\mu,\mu^\prime},
\end{equation}
where $\delta_{\mu,\mu^\prime}$ denotes the Kronecker delta.

The completion of $\mathcal{F}_g$ in the norm induced by
(\ref{scal}) defines the Hilbert space $\mathcal{H}^g_{kin}= L^2
(\dR_{Bohr}, d\mu_{Bohr})$, where $\dR_{Bohr}$ is the Bohr
compactification of the real line and $d\mu_{Bohr}$ denotes the
Haar measure on it \cite{Ashtekar:2003hd}. $\mathcal{H}^g_{kin}$
is  the kinematical space of the gravitational degrees of freedom.
The kinematical Hilbert space of the scalar field is
$\mathcal{H}^\phi_{kin} = L^2(\dR, d\phi)$, and the operators
corresponding to the elementary variables are
\begin{equation}\label{elem}
(\hat{\phi}\psi)(\phi)= \phi \psi(\phi),~~~~\hat{p}_\phi \psi = -i
\hbar \frac{d}{d\phi}\psi .
\end{equation}
The kinematical Hilbert space of the gravitational field coupled
to the scalar field is  defined to be $\mathcal{H}_{kin}=
\mathcal{H}^g_{kin}\otimes \mathcal{H}^\phi_{kin}$.

\subsection{Dynamics}

The  resolution of the singularity
\cite{Ashtekar:2007em,Ashtekar:2006wn,Ashtekar:2006uz,Ashtekar:2003hd,
Bojowald:2006da} is mainly due to the  peculiar way of defining
the  quantum operator corresponding to $H_g$. Let us consider this
issue in more details.

Using the prescription $\{\cdot,\cdot\}\rightarrow \frac{1}{i
\hbar} [\cdot,\cdot]$ and specific factor ordering of operators,
one obtains from (\ref{hamL}) a quantum operator corresponding to
$H_g^{(\lambda)}$ in the form \cite{Ashtekar:2003hd}
\begin{equation}\label{hamRQ}
\hat{H}^{(\lambda)}_g =  \frac{i\, sgn(p)}{2\pi l_p^2 \gamma^3
\lambda^3} \sum_{ijk} \varepsilon^{ijk}\, Tr
\Big(\hat{h}^{(\lambda)}_i \hat{h}^{(\lambda)}_j
(\hat{h}^{(\lambda)}_i)^{-1} (\hat{h}^{(\lambda)}_j)^{-1}
\hat{h}_k^{(\lambda)}\{(\hat{h}_k^{(\lambda)})^{-1},\hat{V}\}\Big).
\end{equation}
One can show \cite{Ashtekar:2003hd} that (\ref{hamRQ}) can be
rewritten as
\begin{equation}\label{hamK}
\hat{H}^{(\lambda)}_g |\mu\rangle = \frac{3}{8 \pi \gamma^3
\lambda^3 l_p^2}\Big(V_{\mu + \lambda}- V_{\mu - \lambda}
\Big)\big(|\mu + 4 \lambda\rangle - 2 |\mu\rangle + |\mu - 4
\lambda\rangle\big),
\end{equation}
where $|\mu\rangle$ is an eigenstate of $\hat{p}$ defined by
(\ref{relB}), and where $V_\mu$ is an eigenvalue of the volume
operator corresponding to $V= |p|^{3/2}$ which  reads
\begin{equation}\label{vol}
    \hat{V}|\mu\rangle = \Big( \frac{4 \pi \gamma |\mu|}{3}\Big)^{3/2}
    l_p^3\; |\mu\rangle =: V_\mu\,|\mu\rangle .
\end{equation}
The quantum operator corresponding to $H_g$ is defined to be
\cite{Ashtekar:2003hd,Ashtekar:2006uz}
\begin{equation}\label{hamQ}
    \hat{H}_g := \hat{H}^{(\lambda)}_g \mid_{\lambda = \mu_o}
,~~~\mbox{~where}~~~0 < \mu_o \in \dR .
\end{equation}
Comparing (\ref{hamQ}) with (\ref{hamR}), and taking into account
(\ref{cur}) we can see that the area of the square $\Box_{ij}$ is
not shrunk to {\it zero}, as required in the definition of the
classical curvature (\ref{cur}), but determined at the {\it
finite} value of the area.

The mathematical justification proposed in
\cite{Ashtekar:2003hd,Ashtekar:2006uz} for such regularization  is
that one cannot define the {\it local} operator corresponding to
the curvature $F^k_{ab}$ because the 1-parameter group
$\hat{N}_\lambda$ is not weakly continuous at $\lambda =0$ in
$\mathcal{F}_g$ (dense subspace of $\mathcal{H}^g_{kin}$). Thus,
the limit $\lambda\,\rightarrow\, 0$ of $\hat{H}^{(\lambda)}_g$
does not exist.  To determine $\mu_o$ one proposes in
\cite{Ashtekar:2003hd,Ashtekar:2006uz,Ashtekar:2006wn} the
procedure which is equivalent to the following: We find  that the
area of the face of the cell $\mathcal{V}$ orthogonal to specific
direction is $Ar = |p|$. Thus the eigenvalue problem for the
corresponding {\it kinematical} operator of an area
$\widehat{Ar}:= |\hat{p}|$, due to (\ref{relB}), reads
\begin{equation}\label{area}
  \widehat{Ar}\,|\mu\rangle = \frac{4\pi \gamma l^2_p}{3}\,
  |\mu| \,|\mu\rangle =: ar (\mu)\,|\mu\rangle,~~~~\mu\in\dR ,
\end{equation}
where $ar (\mu)$ denotes the eigenvalue of $\widehat{Ar}$
corresponding to the eigenstate $|\mu\rangle$. On the other hand,
it is known that in LQG the {\it kinematical} area operator has
{\it discrete} eigenvalues \cite{Ashtekar:1996eg,Rovelli:1994ge}
and the smallest nonzero one, called an area gap $\Delta$, is
given by $\Delta = 2\sqrt{3}\,\pi \gamma l^2_p$. To identify
$\mu_o$ one postulates in \cite{Ashtekar:2006uz} that $\mu_o$ is
such that $ar (\mu_o) = \Delta$, which leads to $\mu_o =
3\sqrt{3}/2$. It is argued
\cite{Ashtekar:2003hd,Ashtekar:2006uz,Ashtekar:2006wn,Ashtekar:2007em}
that one cannot squeeze a surface to the zero value due to the
existence in the universe of the {\it minimum} quantum of area.
This completes the justification for the choice of the expression
defining the quantum Hamiltonian (\ref{hamQ}) offered by LQC.

It is interesting to notice that for the model considered here
(defined on one-dimensional constant lattice) the existence of the
minimum area leads to the reduction of the non-separable space
$\mathcal{F}_g$ to its {\it separable} subspace. It is so because
due to (\ref{relC}) we have
\begin{equation}\label{action}
\hat{N}_{\mu_o}  \,|\mu\rangle =  |\mu + \mu_o\rangle,
\end{equation}
which means that the action of this operator does not lead outside
of the space spanned by $\{|\mu +k\,\mu_o\rangle, \,k\in \dZ\}$,
where $\mu \in \dR$ is fixed.

Finally, one can show (see, e.g.
\cite{Ashtekar:2003hd,Ashtekar:2006uz}) that the equation for
quantum dynamics, corresponding to (\ref{ker}), reads
\begin{equation}\label{quanD}
B(\mu)\;\partial^2_\phi \psi(\mu,\phi) - C^+ (\mu)\psi (\mu +4
\mu_o , \phi) - C^- (\mu) \psi (\mu - 4 \mu_o, \phi) - C^0 (\mu)
\psi(\mu,\phi) = 0,
\end{equation}
where
\begin{equation}\label{BC}
B(\mu):= \Big( \frac{2}{3 \mu_o}\Big)^6 \left[ |\mu + \mu_o|^{3/4}
- |\mu - \mu_o|^{3/4}\right]^6, \quad C^0 (\mu) := - C^+ (\mu) - C^-
(\mu),
\end{equation}
\begin{equation}\label{CC}
C^+ (\mu):= \frac{\pi G}{9 |\mu_o|^3}\left|\;|\mu +3 \mu_o |^{3/2}-
|\mu +\mu_o |^{3/2}\right|, \quad C^-(\mu) := C^+ (\mu - 4 \mu_o ).
\end{equation}

Equation (\ref{quanD}) has been derived formally by making use of
states which belong to $\mathcal{F}:= \mathcal{F}_g \otimes
\mathcal{F}_\phi$, where $\mathcal{F}_g$ and $\mathcal{F}_\phi$
are  dense subspaces of the kinematical Hilbert spaces
$\mathcal{H}^g_{kin}$ and $\mathcal{H}^\phi_{kin}$, respectively.
The space $\mathcal{F}$ provides an arena for the  derivation of
quantum dynamics. However, the {\it physical} states are expected
to be in $\mathcal{F}^\star $, the algebraic dual of $\mathcal{F}$
(see, e.g. \cite{Ashtekar:2003hd,Ashtekar:2006uz} and references
therein). It is known that $\mathcal{F}\subset \mathcal{H}_{kin}
\subset \mathcal{F}^\star $. Physical states are expected to have
the form $<\Psi|:= \sum_\mu \psi(\mu,\phi)<\mu|$, where $<\mu|$ is
the eigenbras of $\hat{p}$. One may give the structure of the
Hilbert space to some subspace of $\mathcal{F}^\star$ (constructed
from solutions to (\ref{quanD})) by making use of the group
averaging method \cite{Marolf:2000iq,Ashtekar:1995zh} and obtain
this way the {\it physical} Hilbert space $\mathcal{H}_{phys}$.

The singularity resolution refers, first of all,  to the behavior
of the expectation value of the matter density operator. Numerical
calculations have shown \cite{Ashtekar:2006wn} that the mean value
of this operator is bounded from above on the states (vectors of
the physical Hilbert space) which are semi-classical
asymptotically.  It is suggested in \cite{Ashtekar:2007em} that
the bounce may occur for the states which are more general than
semi-classical at late times, which demonstrates robustness of LQC
results. Quantum evolution, described by (\ref{quanD}), is
deterministic across the bounce region.  The universe undergoes a
bounce during the evolution from pre-big-bang epoch to
post-big-bang epoch. These are main highlights of LQC   (see, e.g.
\cite{Ashtekar:2008vv} for a complete list).

The argument $\phi$ in $\psi(\mu,\phi)$ is interpreted as an
evolution parameter, $\mu$ is regarded as the physical degree of
freedom. Let us examine the role of the parameter $\mu_o$ in
(\ref{quanD}). First of all, its presence causes that
(\ref{quanD}) is a difference-differential equation so its
solution should be examined on a lattice. It is clear that some
special role must be played by $\mu_o = 0$ as the coefficient
functions of the equation, defined by (\ref{BC}) and (\ref{CC}),
are singular there. One can verify \cite{Ashtekar:2006uz} that as
$\mu_o \rightarrow 0$ the equation (\ref{quanD}) turns into the
Wheeler-DeWitt equation
\begin{equation}\label{wdw}
B(\mu)\;\frac{\partial^2}{\partial\phi^2} \psi(\mu,\phi)- \frac{16
\pi G}{3}\frac{\partial}{\partial\mu}\sqrt{\mu}\frac{\partial
}{\partial\mu}\,\psi(\mu,\phi) =
0,~~~~\mbox{with}~~~B(\mu):=\big|\frac{4\pi\gamma G
\hbar}{3}\,\mu\,\big|^{-3/2} .
\end{equation}

Equation (\ref{quanD}) is not specially sensitive to any other
value of $\mu_o$. Thus, the determination of the numerical value
of this parameter by making use of the mathematical structure of
(\ref{quanD}) seems to be impossible.

\section{Minimum length problem}

The singularity resolution offered by LQC, in the context of flat
FRW universe, is a striking result.  Let us look at the key
ingredients  of the construction of LQC which are responsible for
this long awaited result:

Discussing the mathematical structure of the constraint equation
we have found that $\mu_o$ must be a non-zero  if we wish to deal
with the regular (\ref{quanD}) instead  of the singular
(\ref{wdw}). However, the numerical value of $\mu_o$ cannot be
determined from  the equation (\ref{quanD}). It plays the role of
a {\it free} parameter if it is not specified.

The parameter $\mu_o$ enters the formalism due to the
representation of the curvature of the connection $F^k_{ab}$ via
the holonomy around a {\it loop} (\ref{cur}). The smaller the loop
the better approximation we have. The size of the loop, $\mu_o$,
determines the quantum operator corresponding to the modified
gravitational part of the Hamiltonian (\ref{hamQ}). One may
determine $\mu_o$ by making use of an {\it area} of the loop (used
in fact as a technical tool). Thus, the {\it spectrum} of the
quantum operator corresponding to an area operator,
$\widehat{Ar}$, seems to be a suitable source of information on
the possible values of $\mu_o$. Section III shows explicitly that
the construction of the quantum level  is heavily  based on the
{\it kinematical} ingredients of the formalism. Thus, it is
natural to explore the kinematical $\widehat{Ar}$ of LQC. However,
its spectrum (\ref{area}) is {\it continuous} so it is useless for
the determination of $\mu_o$. On the other hand, the spectrum of
kinematical $\widehat{Ar}$ of LQG is {\it discrete}
\cite{Ashtekar:1996eg,Rovelli:1994ge}. Thus, it was tempting to
use such a spectrum to fix $\mu_o$ postulating that the  minimum
{\it quantum} of area defines the minimum area of the loop
defining  (\ref{hamQ}). This way $\mu_o$ has been fixed.

The physical justification, however, for such procedure is
doubtful because   LQC {\it is not} the cosmological sector of
LQG. The relationship between  LQG and LQC, at the formalisms
level, has been examined recently \cite{Brunnemann:2005in}:  LQC
is a quantization  method {\it inspired} by  LQG (a field theory
with infinitely many degrees of freedom) used to the quantization
of the simplest models of the universe (with finitely many degrees
of freedom) with high symmetries.

The inspiration   consists  mainly in applying the two ingredients
of LQG: (i) modification  of $F^k_{ab}$ by loop geometry, and (ii)
making use of the holonomy-flux algebra. In other words, LQC has
not been {\it derived} from LQG. The construction of LQC has been
carried out by {\it mimicry} of the construction of LQG, but
nothing more. LQG and LQC are two {\it different} quantum models
of two {\it different} systems. Therefore, Eq. (\ref{hamQ})
includes  an insertion by {\it hand} of specific properties of the
spectrum of $\widehat{Ar}$ from LQG into LQC
\cite{Bojowald:2008ik}. After all, the area gap of the spectrum of
$\widehat{Ar}$ of LQG is not a fundamental constant (like the
speed of light, Planck's constant, Newton's constant) so its use
in the context of LQC has poor physical justification.

The singularity problems should be analyzed in terms of the Dirac
observables and physical states \cite{Brunnemann:2005in}. In our
recent papers  we solve the constraints already at the classical
level, make the identification of the Dirac observables and find
the physical phase space before the quantization process. Our
non-standard LQC  is complementary to the Dirac quantization
method which underlies standard LQC. We have found that the energy
density operator has a {\it continuous} bounded spectrum
\cite{Malkiewicz:2009zd}. The  {\it volume} operator has a {\it
discrete} spectrum  bounded from below \cite{Malkiewicz:2009qv}. A
{\it quantum} of the volume is parameterized by the minimum
length.

\section{Conclusions}

It is claimed (see, e.g.
\cite{Ashtekar:2006uz,Ashtekar:2006wn,Ashtekar:2007em}) that the
introduction of the  quantum of area at the kinematical level of
LQC has sound theoretical justification.  We believe we have shown
that it is an {\it ad hoc} assumption without  physical
justification (see \cite{Bojowald:2008ik} for another criticism of
this assumption). Thus, the energy scale characteristic to the Big
Bounce is  unknown. Claiming  that the Planck scale appears
naturally in LQC is still illusive, in spite of the enthusiasm
invoked by the LQC results.

An identification of the {\it energy scale} specific to the Big
Bounce transition is a fundamental problem since it is supposed to
be the energy scale for the unification of gravity with quantum
physics.

The LQC calculations, done for flat FRW model with massless scalar
field, have shown that making an assumption on the existence of a
minimum fundamental length in quantum geometry one can impose
quantum rules onto the expression for the classical constraint
(Hamiltonian) in such a way that some solutions to the equation
describing the evolution of the universe lead to finite
expectation value for the matter density at any value of the
evolution parameter. It is an interesting result which
demonstrates the powerfulness of LQC. However,  further
investigations is needed for finding solution to the  {\it minimum
length} problem. We suggest that the solution may come from {\it
observational} cosmology. For instance, an identification of the
microscale specific to a {\it foamy} structure of space would be
helpful.

\begin{acknowledgments}
We are grateful to Tomasz Paw{\l}owski and {\L}ukasz Szulc for
helpful discussions.
\end{acknowledgments}


\begin{thebibliography}{99}

\bibitem{MTW} C. W. Misner, K. S. Thorne and J. A. Wheeler \textit{Gravitation}
(San Francisco: W. H. Freeman and Company, 1973).

\bibitem{SWH} S. W. Hawking and  G. F. R. Ellis \textit{The large scale structure
of space-time} (Cambridge, Cambridge University Press, 1975).

\bibitem{JPAK} J. Pleba\'{n}ski and A. Krasi\'{n}ski \textit{An Introduction to
General Relativity and Cosmology} (Cambridge, Cambridge University
Press, 2006).

\bibitem{JMMS} J.~M.~M.~Senovilla,
``Singularity Theorems and their Consequences'', Gen. Rel. Grav.
{\bf 30} (1998) 701.

\bibitem{Ashtekar:2003hd}
  A.~Ashtekar, M.~Bojowald and J.~Lewandowski,
  ``Mathematical structure of loop quantum cosmology'',
  Adv.\ Theor.\ Math.\ Phys.\  {\bf 7} (2003) 233
  [arXiv:gr-qc/0304074].

\bibitem{Ashtekar:2006uz}
  A.~Ashtekar, T.~Paw{\l}owski and P.~Singh,
  ``Quantum nature of the big bang: An analytical and numerical
  investigation'',
  Phys.\ Rev.\  D {\bf 73} (2006) 124038
  [arXiv:gr-qc/0604013].

\bibitem{Ashtekar:2006wn}
  A.~Ashtekar, T.~Paw{\l}owski and P.~Singh,
  ``Quantum nature of the big bang: Improved dynamics'',
  Phys.\ Rev.\  D {\bf 74} (2006) 084003
  [arXiv:gr-qc/0607039].

\bibitem{Ashtekar:2007em}
  A.~Ashtekar, A.~Corichi and P.~Singh,
  ``On the robustness of key features of loop quantum cosmology'',
  Phys.\ Rev.\  D {\bf 77} (2008) 024046
  [arXiv:0710.3565 [gr-qc]].

\bibitem{Bojowald:2006da}
  M.~Bojowald,
  ``Loop quantum cosmology'',
  Living Rev.\ Rel.\  {\bf 8} (2005) 11
  [arXiv:gr-qc/0601085].

\bibitem{TT} T. Thiemann \textit{Modern Canonical Quantum General Relativity}
(Cambridge: Cambridge University Press, 2007).

\bibitem{CR} C. Rovelli \textit{Quantum Gravity}
(Cambridge: CUP, 2004).

\bibitem{Ashtekar:2004eh}
  A.~Ashtekar and J.~Lewandowski,
  ``Background independent quantum gravity: A status report'',
  Class.\ Quant.\ Grav.\  {\bf 21} (2004) R53
  [arXiv:gr-qc/0404018].

\bibitem{PAM} P. A. M. Dirac, \textit{Lectures on Quantum Mechanics}
(New York: Belfer Graduate School of Science Monographs Series,
1964).

\bibitem{HT} M. Henneaux and C. Teitelboim, \textit{Quantization of Gauge Systems}
(Princeton: Princeton University Press, 1992).

\bibitem{Ashtekar:1996eg}
  A.~Ashtekar and J.~Lewandowski,
  ``Quantum theory of geometry. I: Area operators'',
  Class.\ Quant.\ Grav.\  {\bf 14} (1997) A55
  [arXiv:gr-qc/9602046].

\bibitem{Rovelli:1994ge}
  C.~Rovelli and L.~Smolin,
  ``Discreteness of area and volume in quantum gravity'',
  Nucl.\ Phys.\  B {\bf 442} (1995) 593
  [Erratum-ibid.\  B {\bf 456} (1995) 753]
  [arXiv:gr-qc/9411005].

\bibitem{Marolf:2000iq}
  D.~Marolf,
  ``Group averaging and refined algebraic quantization: Where are we
  now?'',
  arXiv:gr-qc/0011112.


\bibitem{Ashtekar:1995zh}
  A.~Ashtekar, J.~Lewandowski, D.~Marolf, J.~Mourao and T.~Thiemann,
  ``Quantization of diffeomorphism invariant theories of connections with local
  degrees of freedom'',
  J.\ Math.\ Phys.\  {\bf 36} (1995) 6456
  [arXiv:gr-qc/9504018].

\bibitem{Ashtekar:2008vv}
  A.~Ashtekar,
  ``Quantum Space-times: Beyond the Continuum of Minkowski and
  Einstein'',
  arXiv:0810.0514 [gr-qc].

\bibitem{Brunnemann:2005in}
  J.~Brunnemann and T.~Thiemann,
  ``On (cosmological) singularity avoidance in loop quantum
  gravity'',
  Class.\ Quant.\ Grav.\  {\bf 23} (2006) 1395
  [arXiv:gr-qc/0505032].

\bibitem{Malkiewicz:2009zd}
  P.~Malkiewicz and W.~Piechocki,
  ``Energy Scale of the Big Bounce,''
   Phys.\ Rev.\  D {\bf 80}, 063506 (2009) [arXiv:0903.4352].

\bibitem{Malkiewicz:2009qv}
  P.~Malkiewicz and W.~Piechocki,
  ``Turning big bang into big bounce: Quantum dynamics,''
  arXiv:0908.4029 [gr-qc].

\bibitem{Bojowald:2008ik}
  M.~Bojowald,
  ``Consistent Loop Quantum Cosmology'',
  Class.\ Quant.\ Grav.\  {\bf 26} (2009) 075020
  [arXiv:0811.4129 [gr-qc]].




\end{thebibliography}
\end{document}